\documentclass{kapproc} 

\usepackage{t1enc}

\usepackage{procps} 

\usepackage[xdvi]{graphicx}

\upperandlowercase

\kluwerbib 

\begin{document}

\articletitle{Kinematics in the Starbursting \\Circumnuclear Region of M100}


\author{Emma L. Allard}
\affil{Centre for Astrophysics Research, University of Hertfordshire\\
Hatfield, Herts, AL10 9AB, UK}
\email{allard@star.herts.ac.uk}

\author{Reynier F. Peletier}
\affil{Kapteyn Astronomical Institute, University of Groningen}

\author{Johan H. Knapen}
\affil{Centre for Astrophysics Research, University of Hertfordshire}

\begin{abstract}
We have obtained integral-field spectroscopic data, using the SAURON instrument, of the bar and starbursting circumnuclear region in the barred spiral galaxy M100. From our data we have derived kinematic maps of the mean velocity and velocity dispersion of the stars and the gas, which we present here. We have also produced maps of the total, [O{\sc iii}], and H$\beta$ intensity. The gas velocity field shows significant kinematic signatures of gas streaming along the inner part of the bar, and of density wave streaming motions across the miniature spiral arms in the nuclear pseudo-ring. The stellar velocity field shows similar non-circular motions. The gas velocity dispersion is notably smaller where the star formation occurs in the nuclear zone and H{\sc ii} regions. 
\end{abstract}


\section{Introduction}
The central regions of spiral galaxies are often found to be harbouring a large amount of star formation (Buta \& Combes 1996; Knapen 2004), which can be organised into a ring a kiloparsec or so in radius. These rings are relatively common; Knapen (2004) found that these nuclear or pseudo-rings can be found in 20\% of all spiral galaxies. Galaxies containing nuclear rings appear to be preferentially barred; a large scale bar has been proposed as a mechanism to transport large amounts of gas inwards. Resonances in the non-axisymmetric potential are set up within the galaxy, most importantly the Inner Lindblad Resonances. The inflowing gas is effectively trapped here, and accumulates into a ring. Through some mechanism (e.g., Elmegreen 1994; Knapen et al. 1995; Ryder et al. 2001) the gas trapped here can undergo massive amounts of star formation.

Theoretical considerations and numerical modelling confirm the dynamical link between large-scale bars and nuclear regions (e.g., Shlosman \& Heller 2002). Kinematic observations are needed, however, to constrain the behaviour of the gas and the stars, and to study how the gas inflow, regulated by the bar, leads to star formation. 

The gas in circumnuclear regions can be observed relatively easily, and in the optical wavelength range there are a number of strong emission lines from which to measure the kinematics. The gas, although an important factor in the dynamics, contains only a fraction of the total mass contained in the region, and will follow the potential created by the stars (and possibly the dark matter). Stellar kinematic observations which can directly trace the stellar orbits are needed to observationally confirm and expand the theoretical picture.

This work focuses on the weakly barred spiral galaxy M100 (NGC 4321), which hosts a circumnuclear, star forming pseudo-ring. 

Using the SAURON integral-field spectrograph (Bacon et al. 2001) we have observed the circumnuclear regions as well as the entire bar, and have derived kinematic maps of both the gas and the stars. We also present maps of the total, [O{\sc iii}] and H{$\beta$} intensity derived from our datacube.

\section{Observations}
The SAURON integral-field spectrograph was used on the 4.2m William Herschel Telescope on La Palma on 2003, May 2. The low resolution mode was used, giving a field of view of 33$\times$41 arcsec, fully sampled by 1431 square lenses, 0.94$\times$0.94 arcsec in size, each producing a spectrum. Another 146 lenses cover a region adjacent to the field for simultaneous observation of the sky background. The wavelength range 4800-5380\AA  ~is covered at 4.2\AA  ~spectral resolution, with a sampling of 1.1\AA ~per pixel. This range contains the stellar absorption features H{$\beta$}, Mg\emph{b}, some Fe{\sc i} features, and the emission lines H{$\beta$}, [O{\sc iii}] and [N{\sc ii}]. To cover the complete bar and circumnuclear region of M100, three pointings were made. Each field was exposed for 3{$\times$}1800s, and dithered by a small number of lenses to avoid systemic errors.
 
\section{Data Reduction}
The data were reduced using the specially developed XSauron software (Bacon et al. 2001). This included bias and dark subtraction, the extraction of the spectra using a fitted mask model, low frequency flat-fielding, cosmic-ray removal, wavelength calibration, sky subtraction, and flux calibration. The spectral resolution was also homogenised across the field, and the wavelength scale truncated to achieve a common range. The reduction of integral-field data results in a datacube, which contains two spatial dimensions and one dispersion dimension.

The exposures were then merged, and the three fields mosaiced together. The final datacube was spatially binned using the Voronoi 2D binning method of Cappellari \& Copin (2003) to achieve a constant S/N across the field. The data was binned to an S/N ratio of 35 per resolution element.

\subsection{Kinematics}
The kinematics were fitted using the Penalized Pixel Fitting (PPxF) method of Cappellari \& Emsellem (2004). This method allows the specification of the pixel range over which to fit the spectra, and thus to exclude strong emission lines. The PPxF method finds the best fit to the galaxy spectrum by convolving a template stellar spectrum with a corresponding line-of-sight velocity distribution (LOSVD). The LOSVD is parameterised by a Gauss-Hermite function, from which we can then obtain the mean velocity and the velocity dispersion. To avoid template mismatch problems a spectral library is used, from which an optimal template is derived. We have used the single-metallicity stellar population models of Vazdekis (1999). The optimal template may then be subtracted from the original galaxy spectrum, producing a pure emission line spectrum from which the gas kinematics can be derived, in our case from the [O{\sc iii}] line. Details about errors in this method can be found in Emsellem et al. (2004). 

\begin{figure}
\centering
  \includegraphics[height=.35\textheight]{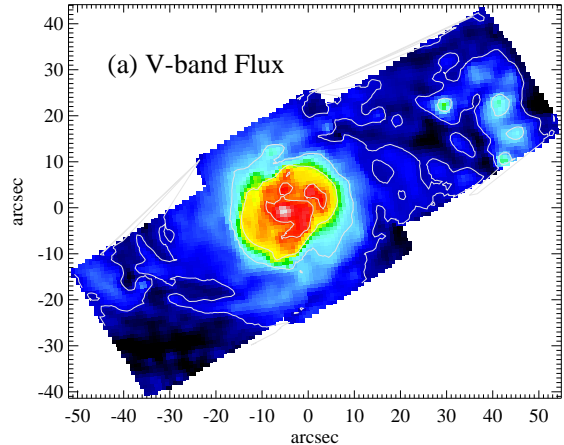}
  \caption{Reconstructed total intensity, showing the nuclear peak, pseudo-ring and the extent of the large bar. North is up and East to the left.}
\end{figure}

\section{Results and Discussion}
The reconstructed total intensity map of M100 (Fig.~1) was produced by collapsing our datacube along the spectral dimension between the full spectral range of 4800-5380{\AA}. We find a bright central source surrounded by spiral armlets which form the pseudo-ring. Two offset dustlanes can be seen entering from the East and from the West. A number of H{\sc ii} regions are present to the NW.
 Figures~2a and 2b were created by collapsing the datacube between the spectral range containing the emission lines [O{\sc iii}] and H{$\beta$} respectively. M100 is classified as a LINER galaxy, and it is not surprising that the  [O{\sc iii}] intensity is the strongest at the centre of the galaxy (Fig.~2a). This is a high-ionization line and suggests an AGN is present.
The H{$\beta$} intensity shows where the star formation is occurring, and is most prominent in the circumnuclear ring and the H{\sc ii} regions. 

\begin{figure}
  \includegraphics[height=.3\textheight]{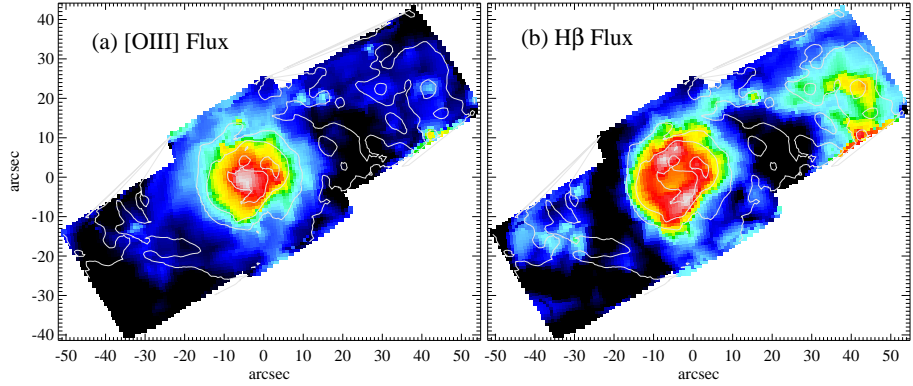}
  \caption{(a). Reconstructed [O{\sc iii}] intensity. (b). Reconstructed H{$\beta$} intensity.}
\end{figure}

The gas velocity field (Fig.~3a) is dominated by circular motion, however an S-shaped deviation is clearly visible in the center of the field. This may be interpreted as the combination of a spiral density wave propagating through the region, and gas streaming along the bar. The gas velocity map here agrees very well with the H{$\alpha$} Fabry-Perot map of Knapen et al. (2001). Spiral density waves are propagating waves of higher and lower stellar density, which lead to non-circular motions of the gas and stars, creating kinks in the velocity field (see also Fathi 2004). 

The stellar velocity field (Fig.~3b) shows predominantly circular motion, although there are distortions present similar to those seen in the gas velocity field. The stellar velocity field does not show any strong streaming along the bar close to the center, as in the case of the gas.

The gas velocity dispersion (Fig.~3c) has a large value at the very center, surrounded by a ring of low dispersion material that coincides with the bright star forming region as seen in the intensity map (Fig.~1a). This suggests that there is a large amount of cold gas present from which stars have recently formed.
The stellar velocity dispersion (Fig.~3d) shows a high region for example where the H{\sc ii} regions are found, in contrast to the gas dispersion. Other areas of large dispersion can be correlated with areas in the gas dispersion map.

\begin{figure}
  \includegraphics[height=.5\textheight]{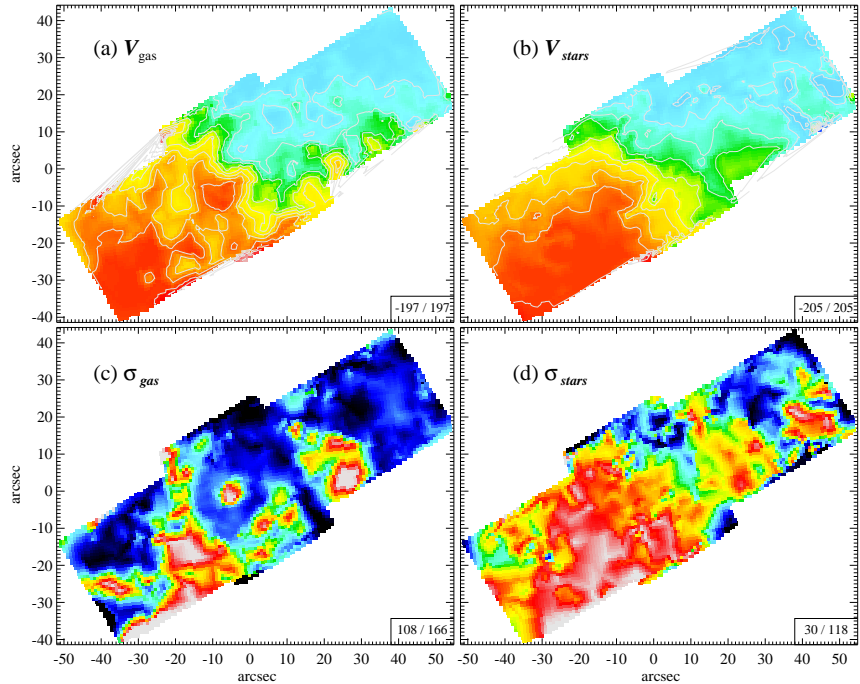}
  \caption{(a). The gas velocity field. (b). The stellar velocity field. (c). The gas velocity dispersion field. (d). The stellar velocity dispersion field. Maximum and minimum values are displayed at the corner of each plot in kms$^{-1}$. Contours are displayed on the velocity fields at intervals of 20 kms$^{-1}$. }
\end{figure}

\section{Conclusions}

We present here our preliminary results of integral-field spectroscopic observations of the circumnuclear and bar regions of M100. Maps of the reconstructed total, [O{\sc iii}], and H{$\beta$} intensity as well as kinematic maps of stellar and gas velocity and velocity dispersion are shown. We confirm the presence of non-circular motions due to a spiral density wave and to streaming along the bar. We observe low gas velocity dispersions where the massive star formation occurs in the nuclear pseudo-ring.

The main purpose of the work done so far is to provide observational constraints on dynamical models of the circumnuclear region of M100. Gas kinematics alone are not enough and through our new data we will be able to obtain more accurate estimates for quantities such as the bar pattern speed and the characteristic stellar bar orbits.  In addition to the kinematical information, the presence of stellar absorption and gas emission lines in our data will allow the study of the stellar populations in the star forming regions, to learn about the detailed physical processes which connect galaxy dynamics with massive star formation under the influence of a bar.

\begin{acknowledgments}
Based on observations obtained at the William Herschel Telescope, operated on the island of La Palma by the Isaac Newton Group in the Spanish Observatorio del Roque de los Muchachos of the Instituto de Astrof\'{i}sica de Canarias.
\end{acknowledgments}

\end{document}